\documentclass[twocolumn,10pt,prd]{revtex4-1}

\def\mysections#1{{\bf #1.} }

\usepackage[dvips]{graphicx}
\usepackage{epsfig,amsmath,amssymb,verbatim,mathrsfs,array,layout,textcomp,amssymb,latexsym, slashed,graphicx,bbm}

\usepackage{graphicx}
\usepackage{amsmath}
\usepackage{amssymb}
\usepackage{appendix}
\usepackage{verbatim,mathrsfs,array,layout,textcomp,latexsym, slashed}

\usepackage{amsmath, amssymb, graphicx, rotating}
\usepackage{hyperref}
\hypersetup{
     colorlinks   = true,
     citecolor    = red,
 linkcolor=blue
}
\usepackage[usenames,dvipsnames]{color}

\usepackage{bm}

\newcommand{\beq}{\begin{eqnarray}}% can be used as {equation} or {eqnarray}
\newcommand{\eeq}{\end{eqnarray}}

\def\beqa{\begin{eqnarray}}
\def\eeqa{\end{eqnarray}}

\newcommand{\bv}{\left(\begin{array}{c}}
\newcommand{\ev}{\end{array}\right)}
\newcommand{\bmtwo}{\left(\begin{array}{cc}}
\newcommand{\bmthree}{\left(\begin{array}{ccc}}
\newcommand{\emn}{\end{array}\right)}
\newcommand{\bmtwoc}{\left\{\begin{array}{cc}}
\newcommand{\bmthreec}{\left\{\begin{array}{ccc}}
\newcommand{\emnc}{\end{array}\right\}}
\newcommand{\ba}{\begin{array}}
\newcommand{\ea}{\end{array}}

\newcommand{\be}{\begin{equation}}
\newcommand{\ee}{\end{equation}}

\def\lsim{\mathrel{\rlap{\lower4pt\hbox{\hskip1pt$\sim$}}
     \raise1pt\hbox{$<$}}}         %less than or approx. symbol
\def\gsim{\mathrel{\rlap{\lower4pt\hbox{\hskip1pt$\sim$}}
     \raise1pt\hbox{$>$}}}         %greater than or approx. symbol

\begin{document}

\font\mini=cmr10 at 0.8pt

\title{
Absorption of light dark matter in semiconductors
}

\author{Yonit Hochberg}\email{yhochberg@lbl.gov}
\author{Tongyan Lin}\email{tongyan@berkeley.edu}
\author{Kathryn M. Zurek}\email{kzurek@berkeley.edu}
\affiliation{Theoretical Physics Group, Lawrence Berkeley National Laboratory, Berkeley, CA 94720 \\ Berkeley Center for Theoretical Physics, University of California, Berkeley, CA 94720}

\date{\today}
\begin{abstract}
    Semiconductors are by now well-established targets for direct detection of MeV to GeV dark matter via scattering off electrons. We show that semiconductor targets can also detect significantly lighter dark matter via an absorption process.  When the dark matter mass is above the band gap of the semiconductor (around an eV), absorption proceeds by excitation of an electron into the conduction band. Below the band gap, multi-phonon excitations enable absorption of dark matter in the 0.01~eV to eV mass range. Energetic dark matter particles
emitted from the sun can also be probed for masses below an eV.
We derive the reach for absorption of a relic kinetically mixed dark photon or pseudoscalar in germanium and silicon, and show that existing direct detection results already probe new parameter space.
With only a moderate exposure, low-threshold semiconductor target experiments can exceed current astrophysical and terrestrial constraints on sub-keV bosonic dark matter.
\vspace{0.2cm}
\end{abstract}

\maketitle

%%%%%%%%%%%%%%%%%%
\section{Introduction}
One of the biggest mysteries in modern physics is the identity of dark matter (DM).  For over three decades, the dominant candidate for DM has been the Weakly Interacting Massive Particle (WIMP), which has served as a guide to theory and experiment. Indeed, existing direct detection experiments have been extremely successful in constraining DM in the GeV to TeV mass range~\cite{Aprile:2012nq,Akerib:2015rjg,Agnese:2014aze,Tan:2016zwf}, and ton-scale future detectors~\cite{Aprile:2015uzo,Akerib:2015cja} will further improve reach into this parameter space.

The exploration of DM beyond the WIMP has gained traction in recent years, both theoretically and experimentally. For sub-GeV DM, various targets have been proposed for detection of DM via scattering processes.
These include electronic ionization~\cite{Essig:2011nj} as well as inelastic photon emission~\cite{Kouvaris:2016afs} in atomic targets, excitation to a conduction band in a semiconductor~\cite{Essig:2011nj,Graham:2012su,Lee:2015qva,Essig:2015cda}, production of scintillation photons~\cite{Derenzo:2016fse}, and ejection of valence electrons in graphene~\cite{Hochberg:2016ntt}, all sensitive in principle to DM of $\sim{\rm MeV}-{\rm GeV}$ mass. (Indeed, constraints on DM scattering with electrons in this mass range using Xenon10 data have already been derived~\cite{Essig:2012yx}.)
The breaking of Cooper pairs in a superconductor~\cite{Hochberg:2015pha,Hochberg:2015fth}, as well as a two-excitation process in superfluid helium~\cite{Schutz:2016tid}, are both sensitive in principle to even lighter DM with mass down to the warm DM limit of $\sim$~keV.

A number of well-motivated bosonic DM candidates can have even lower masses, below a keV~\cite{Essig:2013lka}.  These candidates can be probed in the same systems via an absorption process, where all of the energy of the incoming DM particle is absorbed.
Various mechanisms have been studied in the literature.   Refs.~\cite{Dimopoulos:1985tm,Avignone:1986vm} considered absorption of solar axions in atomic and semiconductor targets, while Refs.~\cite{An:2014twa,Pospelov:2008jk} derived direct detection constraints on relic vector DM via an atomic transition in Xenon, which is sensitive to DM masses above 12~eV. Emission of an athermal phonon also allows DM absorption on a conduction electron in a superconductor~\cite{Hochberg:2016ajh}, probing DM with meV to 10~eV mass.

The purpose of this note is to show that for masses in the (m)eV to keV range, absorption of relic bosonic DM in semiconductor targets such as germanium and silicon is highly competitive and complementary to atomic and superconductor targets.  A number of low-threshold direct direction experiments employ semiconductor targets, with current sensitivity to electronic energy depositions as low as $\sim 50$~eV in CDMSlite~\cite{Agnese:2015nto} and DAMIC~\cite{Aguilar-Arevalo:2016ndq}. In the near future,
such experiments may have thresholds as low as a few~eV, with total exposures up to \mbox{$\sim$~kg-year~\cite{Cushman:2013zza,Essig:2015cda}.}

When the DM energy is above the ($\sim$\,eV) band gap for electron excitations, DM absorption proceeds in semiconductors both through inelastic processes (via direct band transitions) and athermal phonon emission (as in superconductors).
This is possible for halo DM with mass above the band gap as well as for for light DM that is emitted from the sun, which has typical energies of $10-1000$\,eV. The electronic excitations can then be observed either directly in the form of secondary electron-hole pairs, or in the form of phonons via an amplification process.

Importantly, there is another process that potentially allows access to DM with energy below the semiconductor band gap.  Although electron excitations are not allowed, there is no gap for phonon excitations. If two phonon excitations are created, the kinematics of the process change: the excitations are back-to-back, which allows for momentum conservation while all the energy of the DM is absorbed. No electron is excited across the gap, and thus detection of low-energy athermal phonons is crucial. This is the concept proposed in the context of superfluid helium~\cite{Schutz:2016tid}, and which can also be applied to the case of semiconductors below the band gap.

For both electron and multi-phonon excitations, the DM absorption rate can be related to the measured optical properties of the material, providing an excellent description of all relevant processes. In what follows, we will use data to determine the absorption rate for both non-relativistic halo DM and DM emitted from the sun, finding excellent reach for sub-keV bosonic DM with semiconductor targets. In addition, we obtain constraints from existing semiconductor and xenon experiments for DM masses above 50~eV.

\begin{figure*}[t!]
\begin{center}
	\includegraphics[width=0.47\textwidth]{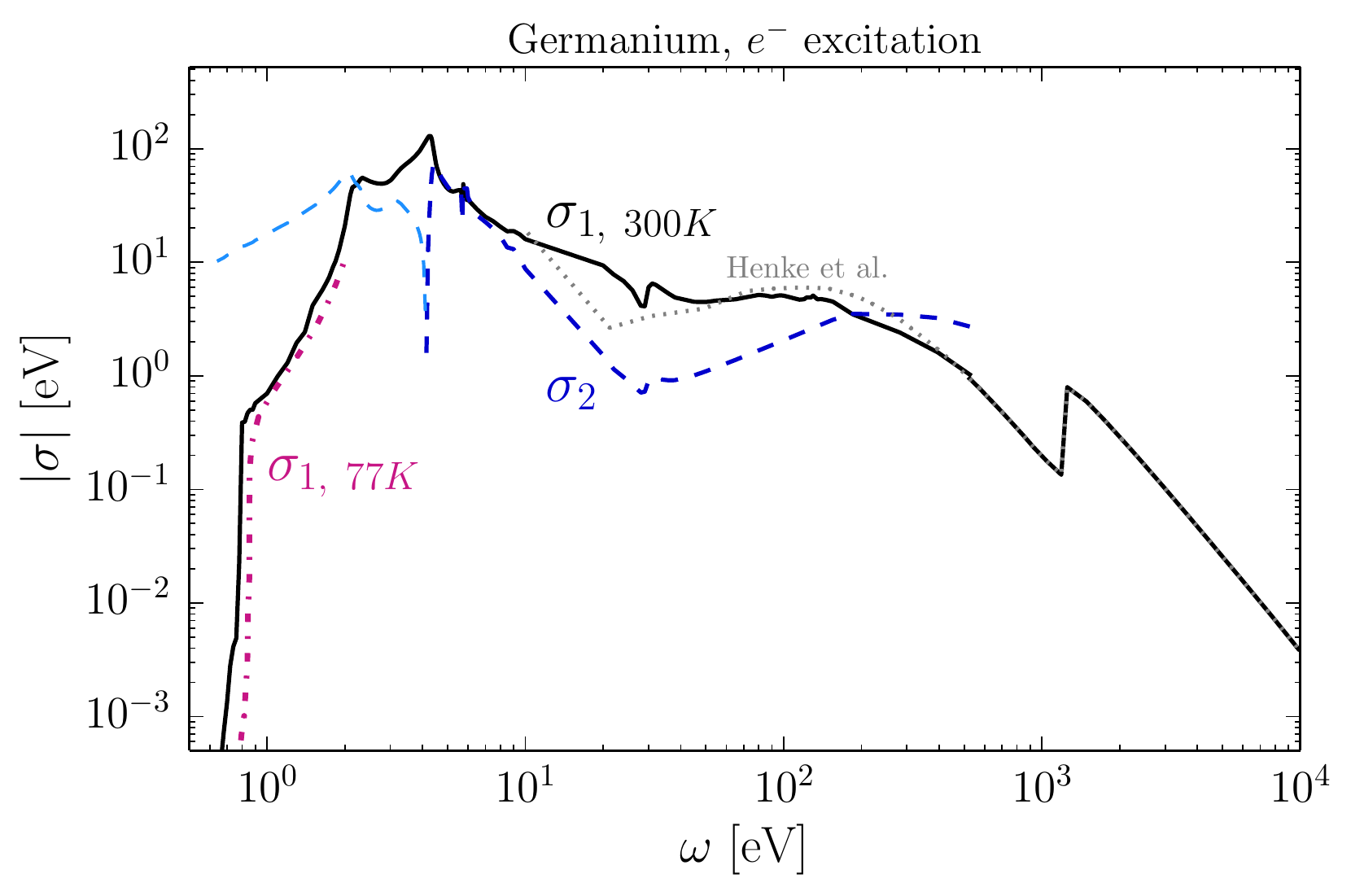}
    \includegraphics[width=0.47\textwidth]{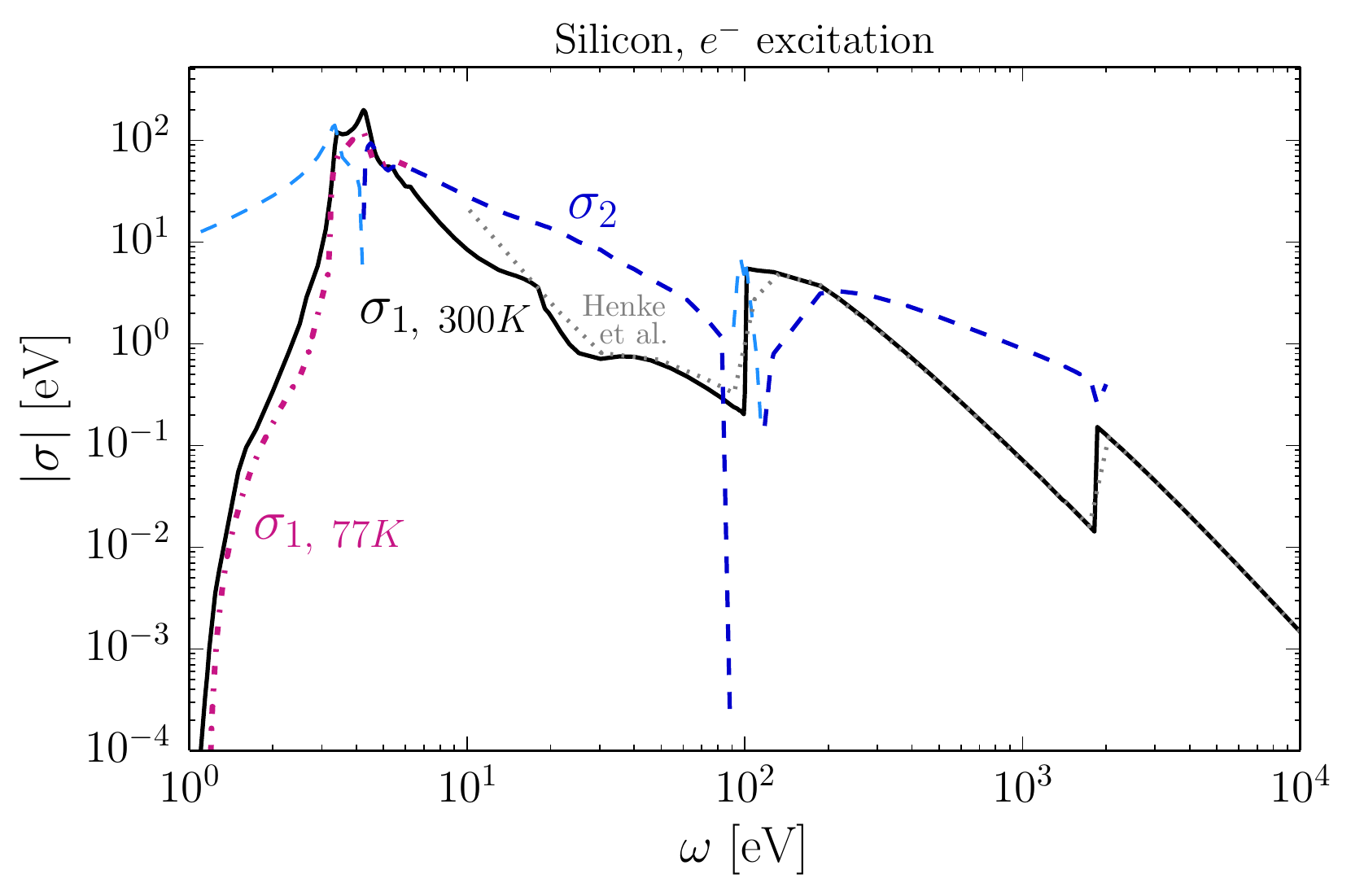}\\
    \includegraphics[width=0.47\textwidth]{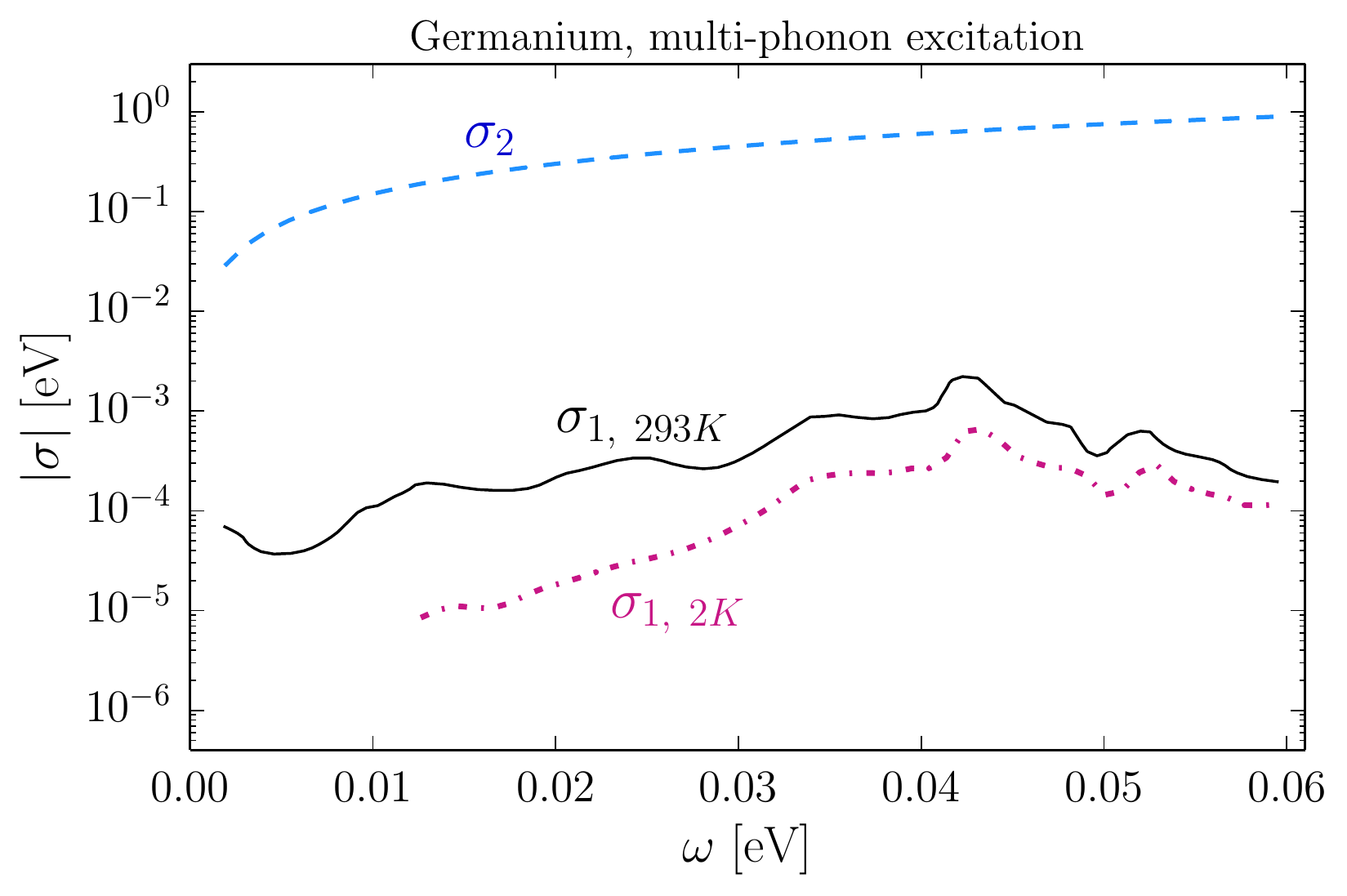}
    \includegraphics[width=0.47\textwidth]{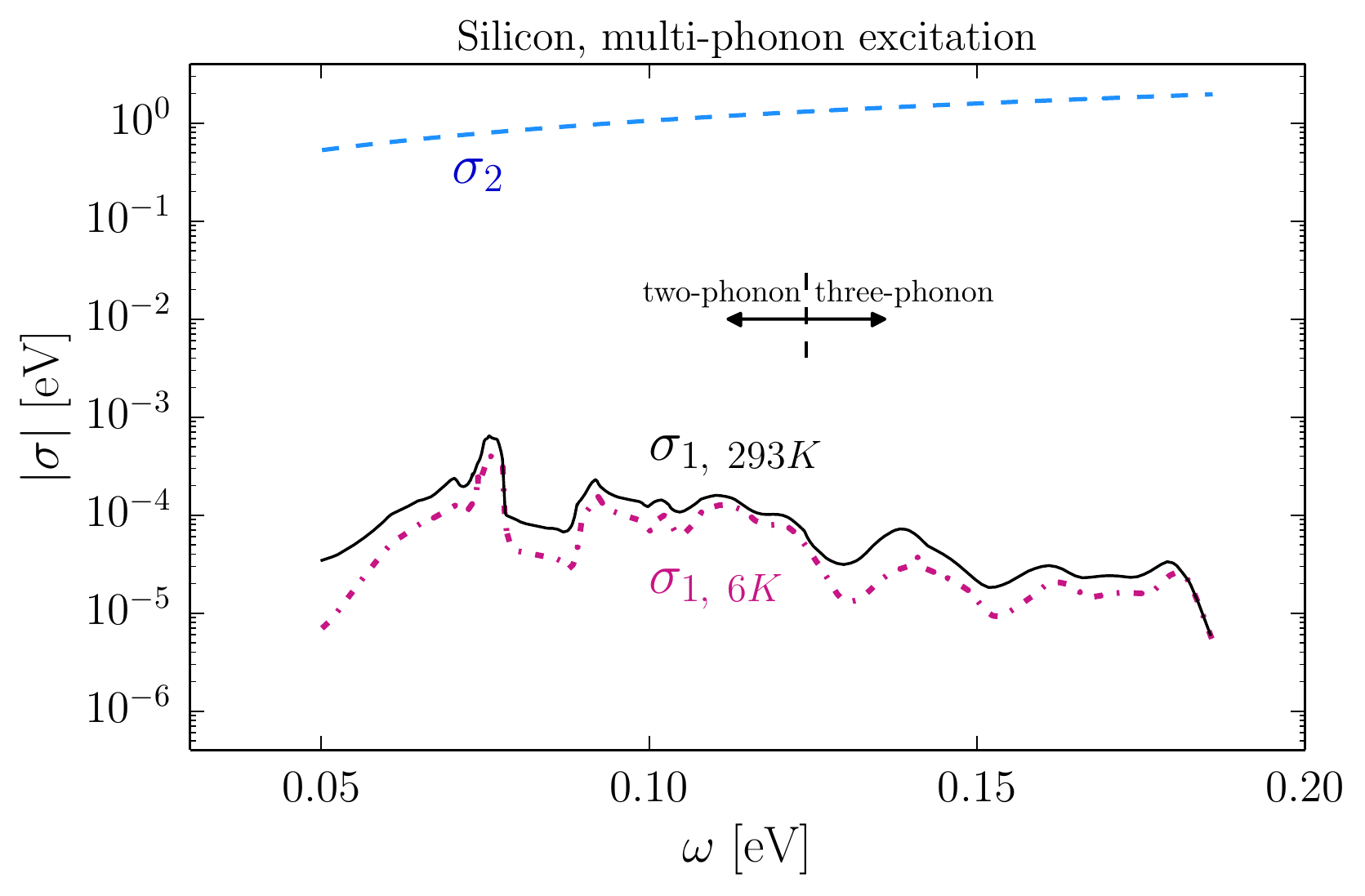}
\end{center}
 \caption{{\bf Top:} Real and imaginary parts of the conductivity in germanium~({\it left}) and silicon~({\it right}) due to electron excitations, using data obtained from Refs.~\cite{opticalconstantsGe} and~\cite{opticalconstantsSi}, respectively. Above $\sim$1 keV, we obtain $\sigma_1$ from Henke et al.~\cite{Henke:1993eda}. The conductivity is shown at room temperatures 290-300 K and is expected to be approximately independent of temperature well above the band gap. We also show the absorption at $T = 77$ K for both silicon~\cite{Holland2003} and germanium~\cite{DashNewman55}. Dashed curves shown in darker (lighter) blue denote positive (negative) values of $\sigma_2$. For comparison, the $\sigma_1$ obtained from the semi-empirical theoretical calculation of Henke et al.~\cite{Henke:1993eda} is shown as the dotted gray line. {\bf Bottom:} Real and imaginary parts of the conductivity in germanium~({\it left}) and silicon~({\it right}) in the regime of multi-phonon excitations, obtained from Refs.~\cite{IkezawaGe} and~\cite{IkezawaSi}, respectively. The measured germanium absorption is primarily due to two-phonon excitation, while the silicon absorption shown includes both two- and three-phonon excitations.
  \label{fig:sig}}
\end{figure*}

%%%%%%%%%%%%%%%%%%
%\vspace{0.3cm}
\section{Method}
The absorption rate of halo DM in a material is given by
\beq\label{eq:R}
R=\frac{1}{\rho}\frac{\rho_{\rm DM}}{m_{\rm DM}}\langle n_e \sigma_{\rm abs}v_{\rm rel}\rangle\,,
\eeq
where $\rho$ is the mass density of the target, $\sigma_{\rm abs}$ is the DM absorption cross section on electrons, $m_{\rm DM}$ is the DM mass, and $\rho_{\rm DM}=0.3\;{\rm GeV}/{\rm cm}^3$ is the local DM mass density.

We will relate the absorption rate of dark photon and pseudoscalar DM to measured optical properties of the target. The absorption rate for photons is determined by the in-medium polarization tensor of the electromagnetic field, $\Pi$, via the optical theorem:
\beq\label{eq:ratephoton}
\langle n_e \sigma_{\rm abs}v_{\rm rel}\rangle_\gamma=-\frac{{\rm Im}\ \Pi(\omega)}{\omega}\,.
\eeq
Here $\omega$ is the energy of the photon that is being absorbed, and we have used the fact that in the local limit, where the three-momentum of the incoming photon $|\vec q|$ can be neglected, the transverse and longitudinal modes of the polarization tensor are of equal size, denoted here by $\Pi(\omega)$. This $\Pi(\omega)$ is related to the complex conductivity $\hat \sigma(\omega) \equiv \sigma_1 +i \sigma_2$, which describes the frequency-dependent response of the system to an EM field,
\beq\label{eq:pisig}
\Pi(\omega)\approx - i \hat \sigma \omega\,.
\eeq
The conductivity is in turn related to the complex index of refraction $\hat n$ by $\hat n^2 = 1 + i \hat \sigma/\omega$.

From Eqs.~\eqref{eq:ratephoton} and~\eqref{eq:pisig} it is clear that the real part of the conductivity $\sigma_1$ is the absorption rate for excitations of energy $\omega$, and
\beq
\langle n_e \sigma_{\rm abs}v_{\rm rel}\rangle_\gamma=\sigma_1\,.
\eeq

For a given target material, one can obtain measurements of the conductivity $\hat \sigma$ at various energies. Then, by relating the absorption rate of DM in the material to that of photons, the sensitivity to a DM candidate can be obtained.
Ref.~\cite{Hochberg:2016ajh} applied this method to an aluminum superconducting target, and obtained the reach for relic kinetically mixed vector and pseudoscalar DM, in the meV to eV mass range.  An analytically derived formula for the absorption rate allowed Ref.~\cite{Hochberg:2016ajh} to extend the constraints to scalar DM.  In Ref.~\cite{An:2014twa}, absorption of kinetically mixed vectors in liquid Xenon was considered, where ionization limits the reach for non-relativistic DM to masses above $\sim 12$~eV. Since the bandgap of a semiconductor such as germanium is smaller (of order $\sim 0.7\;{\rm eV}$), lower DM masses can be probed by absorption in semiconductors through an electron excitation process.

In Fig.~\ref{fig:sig} we show measurements of the optical conductivity in germanium~({\it left}) and silicon~({\it right}), which we will use to obtain the sensitivity of these materials to light DM absorption.  While the absorption rate scales only with $\sigma_1$, the effective coupling of a kinetically mixed dark photon will also depend on $\sigma_2$, so we show both quantities. 

In the top row, we consider the energy range where absorption via electron excitation to the conduction band is relevant, namely above the semiconductor band gap ($0.7$ eV in germanium; 1.1 eV in silicon). For a broad range of energies, the conductivity is obtained using room temperature data from Refs.~\cite{opticalconstantsGe} (germanium) and ~\cite{opticalconstantsSi} (silicon). Note that while error bars are not provided for these measurements, the variation of the measurements between different experiments can be taken as a proxy for the uncertainty. This uncertainty is typically less than a few percent and up to $\sim10\%$ for some energies. Finally, for energies above $\sim$1 keV, $\sigma_1$ is obtained from the semi-empirical theoretical calculations of Henke et al.~\cite{Henke:1993eda}; $\sigma_2$ will not be needed at these energies (see discussion below Eq.~\eqref{eq:kappaeff}).

Since both germanium and silicon are indirect gap semiconductors, absorption at energies near the band gap is phonon-assisted: the electron excitation requires either the emission or absorption of a phonon to conserve energy and momentum. As a result, the presence of thermal phonons can have a substantial effect on the absorption rate. For $\omega \lesssim$ few eV, we also show the absorption at $T=77$~K~\cite{Holland2003,DashNewman55} for comparison with the $T\approx300$~K data. (The difference between absorption at temperatures of 77~K and 4.2~K is even smaller~\cite{MacfarlaneSi,MacfarlaneGe}.) Anticipating the low operating temperatures of SuperCDMS and DAMIC, we use the $T=77$~K data where available.

 At larger energies ($\gtrsim 0.9$~eV in germanium; $\gtrsim 2.5$~eV in silicon), direct band transitions without phonons are possible, and the temperature dependence is expected to be mild.
Similarly, $\sigma_2$ is primarily determined by the band structure and also has only a mild temperature dependence~\cite{DresselGruner}. We thus expect that using the room temperature data is a very good approximation for the low temperature conductivity at these higher energies.

Turning now to energies below the semiconductor band gap, we find that DM absorption can proceed instead by multi-phonon excitations. The process is analogous to infrared photon absorption, which arises due to a second-order coupling of the crystal dipole moment with phonons~\cite{Kress1968,Twophonon2004}.
Here an incoming photon can excite a dipole moment in the lattice by creating phonons, quantized lattice displacements. For germanium and silicon, at least two phonons must be created in this process due to the symmetry of the crystal.

The absorption of photons into multi-phonons has been observed in the $\sim$few~meV$-0.1$~eV energy range for germanium and silicon (see, {\it e.g.}, Refs.~\cite{IkezawaSi,IkezawaGe}). The dominant absorption is due to a two-phonon emission process, where the two phonons are back-to-back. Such two-phonon excitations can only occur for deposited energies below $\sim$0.1~eV, since there is a maximum energy for an acoustic phonon in a lattice. This maximum energy is roughly given by the Debye temperature,  0.03~eV/phonon for germanium and 0.06~eV/phonon for silicon. Thus higher-order three-phonon excitations are required for deposited energies above $\sim$0.1~eV, leading to a smaller absorption rate.

The bottom row of Fig.~\ref{fig:sig} shows the conductivity at these low energies.
In this energy range, $\sigma_2$ is given by $\sigma_2 \approx \omega(1 - n^2)$, with constant index of refraction $n \approx 3.4~(4)$ for silicon (germanium)~\cite{Li1980}. For $\sigma_1$, we again show the absorption at low temperature as well as at room temperature for comparison. Due to the importance of thermal phonons for these absorption processes, the temperature dependence is stronger than the case of electron excitations. (For theoretical studies of the temperature dependence of the two-phonon absorption, see Ref.~\cite{Twophonon2004}.)
In deriving the reach for DM absorption, we use the low-temperature data since the expected operating temperature of SuperCDMS SNOLAB is below 0.1~K.

The multi-phonon excitation for germanium is primarily the two-phonon process, which only extends up to $0.06$~eV due to the lower Debye temperature of germanium.
For silicon, two-phonon excitation can occur for energies up to $\sim 0.12$ eV. The regime of three-phonon absorption has also been observed at higher energies, and the approximate boundary between two- and three-phonon absorption is indicated in the figure.

%%%%%%%%%%%%%%%%%%

%\vspace{0.3cm}
\section{Results}
To estimate the reach, we consider a 1~kg-year exposure for both electron and multi-phonon excitations. While experimental sensitivity to eV scale electron excitations may be achieved in the near future, the energy thresholds for phonon detection are much higher. A significant challenge to detecting multi-phonon excitations is lowering the phonon energy thresholds below an eV, which will require exceptional improvement on current technology and control over environmental noise (see {\it e.g.}, Refs.~\cite{Pyle:2015pya,Hochberg:2015fth}). For a simple comparison between different processes, however, we have assumed a single exposure.

In presenting the sensitivity to light dark matter, we incorporate known backgrounds in the limit-setting procedure, and assume that no other backgrounds are present.  In particular, for electron excitations in silicon we include a flat radioactive background of 300/kg/yr/keV~\cite{Agnese:2016cpb}. For electron excitations in germanium, we include cosmogenic backgrounds, which can be substantial for energies of 100-1000 eV~\cite{golwala}, as well as solar neutrino backgrounds below $\sim 10$ eV. In comparing these backgrounds with the mono-energetic signal, 
we assume the energy resolution is similar to that of \mbox{CDMSlite}~\cite{Agnese:2015nto}, where the energy resolution is modeled as $\sigma^2(E) = \sigma_0^2 + \alpha E + \beta E^2$~\cite{Rito}, except that we set $\sigma_0=0$ in our projections to account for the lower energy thresholds. For each $m_A$, the reach is then derived by considering energies within $2\sigma$ of $m_A$.

Coherent nuclear scattering of solar neutrinos may also be a background to the multi-phonon signal we consider. For germanium, these nuclear recoils peak at energies of $0.1-10$~eV, with a rate of order $1-10$/kg/year/eV~\cite{Essig:2011nj,Hochberg:2015fth}. However, considering only energies below $0.1$ eV where the multi-phonon excitations are relevant, the rate is less than 1/kg/year. Similarly, for silicon the rate below $0.2$~eV is much less than 1/kg/year and can be neglected. 

It was also noted recently that coherent photon scattering on nuclei may be an important background at sub-eV energies, with rates as large as tens of events per kg-year in Ge or Si~\cite{Robinson2016}. However, there are several caveats in using these background rates, and so we do not include them in our reach estimate. First, the rate depends on the abundance of radioactive isotopes in the experiment, which may be controllable to some extent. Second, this background was calculated assuming recoils of free nuclei, which is valid for large recoil energies; for energies of order the Debye energy $\sim 0.01-0.1$ eV we expect there to be a suppression in the energy deposition, similar to the largely recoil-free scattering seen in Mossbauer spectroscopy~\cite{mossbauer}.

%%%%%%%%%%%%%%
\begin{figure*}[t!]
\begin{center}
	\includegraphics[width=0.8\textwidth]{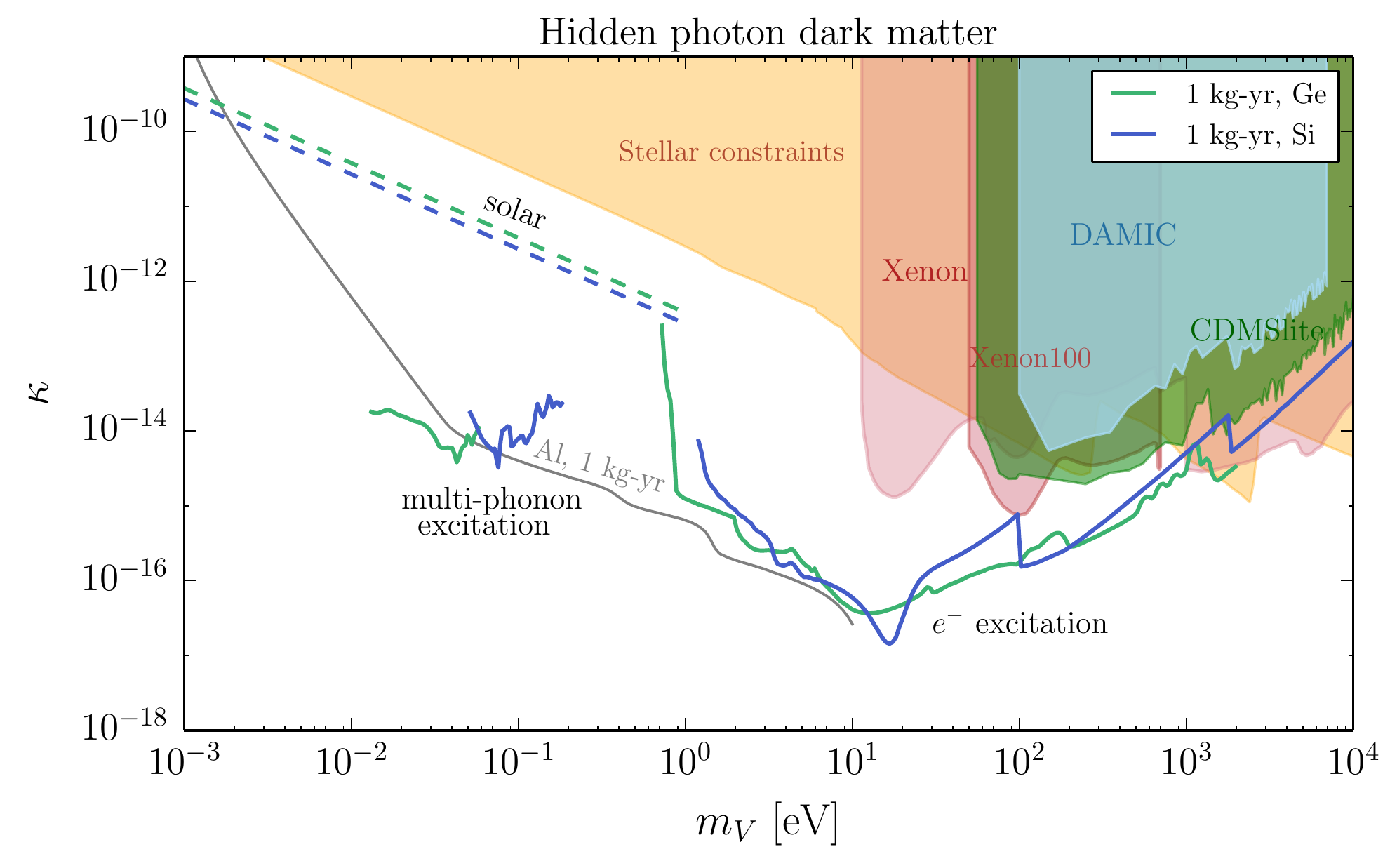}
\end{center}
 \caption{
Estimated reach of a germanium (green lines) and silicon (blue lines) target at 90$\%$ CL  for 1-kg-year exposure, assuming solar neutrino backgrounds only, for absorption of kinetically mixed hidden photon dark matter. For absorption of halo DM (solid lines), we show the reach considering multi-phonon excitations for $m_V = 0.01-0.2$ eV, and  electron excitations for $m_V > 0.6$ eV. The dashed lines show the reach for absorption of dark photons emitted from the sun. Our recast of constraints from CDMSlite (germanium) for $m_V > 56$ eV and DAMIC (silicon) for $m_V > 100$ eV are indicated by the shaded green and blue regions, respectively.  We also show bounds from Xenon10 and Xenon100, including those from Ref.~\cite{An:2014twa} (lighter shaded red) and our own updated Xenon100 limits for 50-700 eV (darker shaded red); the projected reach for 1-kg-year exposure of an aluminum superconducting target (grey line)~\cite{Hochberg:2016ajh}; and stellar emission constraints (shaded orange)~\cite{An:2013b,An:2014twa}.
  \label{fig:hidden}}
\end{figure*}
%%%%%%%%%%%%%%

\subsection{Hidden Photons}

Turning to DM models, we first consider a hidden photon that is kinetically mixed with the hypercharge gauge boson. The hidden photon could compose all of the relic DM, with an abundance set by a misalignment mechanism~\cite{Nelson:2011sf,Arias:2012az,Graham:2015rva}. There is an induced kinetic mixing of the dark photon with the \mbox{photon,}
\beq\label{eq:LHidden}
{\cal L}\supset -\frac{\kappa}{2} F_{\mu\nu}V^{\mu\nu}\,,
\eeq
where $F^{\mu\nu}$ and $V^{\mu\nu}$ are the field strengths for the photon and hidden photon, respectively. A field redefinition of the photon $A_\mu\to A_\mu-\kappa V_\mu$ leads to the canonical basis, where the electromagnetic current picks up a dark charge, $\kappa e V_\mu J_{\rm EM}^{\mu}$ in vacuum.

In-medium effects can substantially alter the polarization tensor $\Pi$, however. For absorption of non-relativistic halo DM, there is an effective mixing angle,
\beq\label{eq:kappaeff}
\kappa_{\rm eff}^2 =\frac{\kappa^2 m_V^4}{\left[m_V^2-{\rm Re}\,\Pi(m_V)\right]^2+\left[{\rm Im}\, \Pi(m_V)\right]^2}\,,
\eeq
where $\Pi$ is related to $\hat \sigma$ ala Eq.~\eqref{eq:pisig}, and the measured conductivities are shown in Fig.~\ref{fig:sig}.  Note that for $m_V \gtrsim 100$~eV, $\kappa_{\rm eff}$ is well-approximated simply by $\kappa$.
The matrix element for absorption of the kinetically mixed hidden photon on electrons is related to that of the photon by $|{\cal M}|^2 = \kappa_{\rm eff}^2 |{\cal M}_\gamma|^2$. Then, the rate in counts per unit time per unit target mass, Eq.~\eqref{eq:R}, is given by
\beq\label{eq:rateHidden}
R=\frac{1}{\rho}\frac{\rho_{\rm DM}}{m_{\rm DM}}\kappa_{\rm eff}^2 \sigma_1(m_V)\,.
\eeq

The projected sensitivity for a hidden photon via absorption in semiconductors at 90\%~CL is shown in Fig.~\ref{fig:hidden} for 1 kg-yr of exposure. For absorption of halo DM, the reach for germanium and silicon  comes from electron excitations for masses above 0.5~eV, while for lower mass it arises from absorption via multi-phonon excitations. Note that for germanium, the dips in sensitivity around 100 eV and 1 keV are due to the electron capture peaks in the cosmogenic background. The projected 90$\%$ CL reach of a superconducting aluminum target in the complementary ${\rm meV}-{\rm eV}$ mass range is depicted as well, for the same exposure~\cite{Hochberg:2016ajh}. As is evident, the two-phonon process provides a powerful probe of bosonic DM in the ${\cal O}(1-100)$~meV mass range, comparable to that of superconductors.

For DM mass below the band gap, we also consider the reach for absorption of hidden photons emitted from the sun.  For $m_V \ll$ eV, the dominant solar production mode for hidden photons is in the longitudinal modes, with a flux that peaks at  $\omega \approx 10-100$~eV. The emitted particles can then be absorbed via electron excitations, with a differential absorption rate given by~\cite{An:2013b}
\begin{align}
	\frac{dR}{d\omega} = \frac{1}{\rho} \frac{d\Phi}{d \omega} \frac{\kappa^2 m_V^2 \sigma_1(\omega)}{\left[ \omega - \sigma_2(\omega) \right]^2 + \left[ \sigma_1(\omega) \right]^2} \, ,
\end{align}
where the flux of hidden photons at the earth is \mbox{$\frac{d\Phi}{d \omega} \propto \kappa^2 m_V^2$.} We integrate this rate over the energy range $1-1000$~eV, following Ref.~\cite{Redondo:2013lna} for the flux, and obtain the reach shown as the dashed lines in Fig.~\ref{fig:hidden}. Due to the strong $\kappa^4$ dependence of the signal, this reach is relatively weak compared to existing constraints.
We note that a ton-scale xenon experiment can achieve a similar sensitivity to semiconductors only if the electronic energy threshold of the former can be lowered to $\sim$100~eV.

%\subsubsection{Current constraints}

Existing limits on absorption of halo DM from Xenon10 and Xenon100 data are shown in Fig.~\ref{fig:hidden} for masses above the ionization threshold in xenon of 12~eV. We include constraints obtained from Ref.~\cite{An:2014twa}, which used 15 kg-day of Xenon10 data~\cite{Angle:2011th} for $m_V =$ 12~eV$-$1~keV and the Xenon100 solar axion search~\cite{Aprile:2014eoa} for \mbox{$m_V>$\,1 keV}. In addition, we have recast the recent Xenon100 low-threshold analysis~\cite{Aprile:2016wwo}, which had a total exposure of 30 kg-year, to obtain updated limits in the mass range $50-700$ eV. Ref.~\cite{Aprile:2016wwo} provides their data in the form of observed photoelectrons (PE) for each event. For a deposited energy of $m_V$, we obtain the distribution in PE using Refs.~\cite{Aprile:2013blg,Shutt:2006ed}, which gives a signal peaked at $(m_V/13.5 \textrm{eV}) \times 20$ PE and with a width of $\sigma \approx \sqrt{m_V/13.5 \textrm{eV}} \times 7$ PE.   Accounting for the experimental efficiency, we compare the signal with the observed counts in a bin of size $4\sigma$ to obtain the 90$\%$~CL limit. Our result  is roughly an order of magnitude stronger than the Xenon10 limit from Ref.~\cite{An:2014twa}, and is shown as the dark red shaded region in Fig.~\ref{fig:hidden}.

%%%%%%%%%%%%%%
\begin{figure*}[th!]
\begin{center}
	\includegraphics[width=0.68\textwidth]{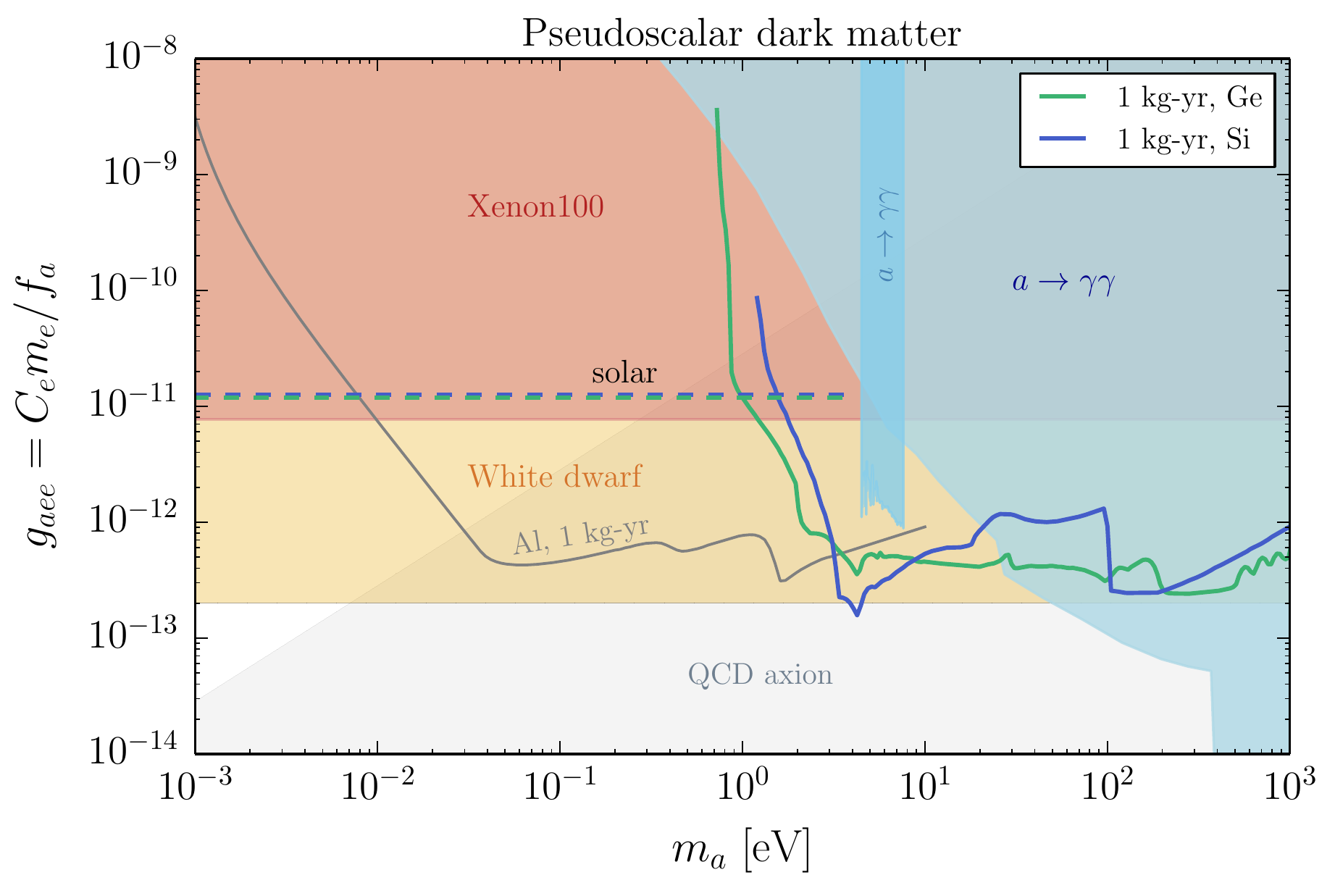}
\end{center}
 \caption{Estimated reach of a germanium (green lines) and silicon (blue lines) target at 90$\%$ CL with a 1-kg-year exposure, assuming solar neutrino backgrounds only, for absorption of pseudoscalar dark matter.  The solid lines show the reach for absorption of halo DM, while the dashed lines are for absorption of pseudoscalars emitted from the sun. The reach of an aluminum superconducting target is given by the solid grey line~\cite{Hochberg:2016ajh}. We show constraints from absorption of solar axions in Xenon100 (shaded red)~\cite{Aprile:2014eoa} and stellar emission from white dwarfs (shaded orange)~\cite{Raffelt:2006cw}.  The range of couplings for the QCD axion is indicated by the shaded grey region.
Constraints on pseudoscalar decays into photons (shaded blue) assume the coupling in Eq.~\eqref{eq:axion_photon}, and come from a line search for $m_a = 4.5-7.5$ eV~\cite{Grin:2006aw} as well as from the extragalactic background light, early reionization, and x-rays~\cite{Arias:2012az}.
  \label{fig:axion}}
\end{figure*}
%%%%%%%%%%%%%%

For comparison, we demonstrate that existing semiconductor targets already start to probe new parameter space for DM mass down to 100~eV. Re-interpreting recent results from \mbox{CDMSlite}~\cite{Agnese:2015nto}, with 70~kg-day exposure on germanium, and DAMIC~\cite{Aguilar-Arevalo:2016ndq}, with 0.6~kg-day exposure on silicon, we obtain limits on absorption of DM in the halo, shown as the shaded green and blue regions in Fig.~\ref{fig:hidden}.

For DAMIC, we derive 90$\%$~CL limits by comparing the DM signal with the observed counts in a single energy bin of width 100~eV. For the mono-energetic absorption signal, we apply the given experimental efficiency and also account for the finite energy resolution of the experiment.   Following Ref.~\cite{Aguilar-Arevalo:2016ndq}, we model the energy resolution by a Fano model, $\sigma^2(E) = \sigma_0^2 + (3.77 \textrm{eV}) FE$ with $F = 0.133 \pm 0.005$. With typical total energy resolution of $\sim 50$ eV, this introduces an additional ${\cal O}(1)$ efficiency for the DM signal to fall in a single bin. Assuming the best-fit background of $\approx$0.5~events/bin, we then obtain upper limits of the DM signal following Ref.~\cite{Feldman:1997qc}, as depicted in Fig.~\ref{fig:hidden}.

We follow a similar procedure to obtain 90$\%$~CL limits from CDMSlite. Here we model the energy resolution with a modified Fano model~\cite{Rito}, given by $\sigma^2(E) = \sigma_0^2 +  \alpha E + \beta E^2$. We fit these constants to the measured energy resolutions given in Table~I of Ref.~\cite{Agnese:2015nto} and include an extra data point for the baseline energy resolution, $\sigma^2(0) = (14\, \textrm{eV})^2$. For $m_V > 100$~eV, we then set conservative limits using the observed counts within single 100~eV bins, making no assumption for the background model. For DM masses closer to the experimental threshold, $m_V = 56-100$ eV, we instead use the 90$\%$~CL upper limit on the rate in the lowest energy bin from Table I of Ref.~\cite{Agnese:2015nto}. Our result is shown in Fig.~\ref{fig:hidden}

In Fig.~\ref{fig:hidden}, we also show existing Xenon10 limits on absorption of solar hidden photons, along with other stellar cooling constraints from the sun, horizontal branch stars, and red giant stars, assuming the dark photon obtains its mass via the Stuckelberg mechanism~\cite{An:2014twa}. (For stellar constraints in the case that the dark photon mass arises from a dark Higgs mechanism, see Ref.~\cite{An:2013b}.)

We learn that semiconductor targets, such as germanium and silicon, are powerful probes of hidden photon DM with mass in the meV$-$keV range, finding a reach that can supersede all existing terrestrial and astrophysical bounds, with only mild exposure.

%\vspace{0.2cm}
\subsection{Pseudoscalars}

Next, we consider a pseudoscalar $a$ that couples to electrons:
\begin{equation}\label{eq:Laxion}
	{\cal L}\supset \frac{g_{aee}}{2 m_e} (\partial_\mu a)\bar e \gamma^\mu \gamma^5 e\,.
\end{equation}
This pseudoscalar may be an axion-like particle, see for example Ref.~\cite{Arias:2012az}.
For comparison, we will show the relation between the mass $m_a$ and coupling constant for the QCD axion in our results:  then the effective coupling can be written as $g_{aee} = C_e m_e/f_a$, with $(0.60\ {\rm meV}/ m_a) = (f_a/10^{10}\ {\rm GeV} )$, and we take $C_e=1/3$ as an upper bound.

 For non-relativistic halo DM, the leading matrix-element-squared for absorption of the pseudoscalar is related to photon absorption by $|{\cal M}|^2  \approx  3 ( g_{aee}/2 m_e)^2 (m_a/e)^2 |{\cal M}_\gamma|^2$~\cite{Pospelov:2008jk,Hochberg:2016ajh}. Then the rate for pseudoscalar absorption is related to the measured conductivity by
\begin{equation}
\label{eq:rateAxion}
	R \simeq \frac{1}{\rho} \frac{\rho_{\rm DM}}{m_{\rm DM}}\frac{3 m_a^2}{4 m_e^2}  \frac{g_{aee}^2}{e^2}   \sigma_1(m_a) \,.
\end{equation}

The expected 90$\%$ CL reach for pseudoscalar DM is shown in Fig.~\ref{fig:axion}, for germanium and silicon targets with 1 kg-year exposure. Here we consider only the reach from electron excitations, in accord with the electron coupling of Eq.~\eqref{eq:Laxion}.
Absorption from multi-phonon excitations is not included, since this process relies on an effective coupling of the DM with the ion displacements in the crystal.  We also depict the projected reach from absorption of halo DM in a superconducting aluminum target~\cite{Hochberg:2016ajh}.

For $m_a \ll$ keV, pseudoscalars can be emitted from the sun and absorbed in the target, with a differential rate given by~\cite{Pospelov:2008jk}
\begin{align}
	\frac{dR}{d\omega} = \frac{1}{\rho} \frac{d\Phi}{d \omega} \frac{ \omega^2}{2 m_e^2}  \frac{g_{aee}^2}{e^2}   \sigma_1(\omega) \, ,
\end{align}
where $d\Phi/d\omega \propto g_{aee}^2$ is the solar flux at the earth. Following Ref.~\cite{Redondo:2013wwa} for the solar flux, we again integrate over the energy range $1-1000$ eV to obtain a reach with semiconductors, shown as the dashed lines in Fig.~\ref{fig:axion}.
In contrast to the hidden photons emitted from the sun, the pseudoscalar flux peaks at around $\omega \approx$ keV. Consequently, we find that experiments with higher energy threshold but larger exposure perform better than low-threshold semiconductor targets. Fig.~\ref{fig:axion}  shows the existing Xenon100 constraint~\cite{Aprile:2014eoa} on DM emission from the sun, which is better than the semiconductor reach. Furthermore, a xenon experiment with a ton-year exposure and keV energy threshold could probe $g_{aee} > 10^{-12}$~\cite{Arisaka:2012pb} from solar emission, similar to the reach of the superconducting target for halo DM.

%\subsubsection{Current constraints}

The strongest constraints for $m_a < {\rm 10\;eV}$ arise from stellar emission of light pseudoscalars in electron-dense environments, such as white dwarfs~\cite{Raffelt:2006cw}. We note that the white dwarf constraint has a factor of few uncertainty, with some of the data actually in favor of the presence of a new particle~\cite{Isern:2008nt,Bertolami:2014wua,Giannotti:2015kwo}.

The pseudoscalar coupling to electrons gives rise to a loop-induced coupling to photons,
\begin{equation}
	\frac{\alpha}{8\pi } \frac{g_{aee}}{m_e} a F_{\mu \nu} \tilde F^{\mu \nu}\,.
	\label{eq:axion_photon}
\end{equation}
This coupling can be modified by ${\cal O}(1)$ effects if the pseudoscalar couples to other charged particles.
Assuming the induced coupling above, we show the constraints on $g_{aee}$ from $a \to \gamma \gamma$ decay, including a line search for $m_a = 4.5-7.5$ eV~\cite{Grin:2006aw}, and the effect of $a \to \gamma \gamma$ on the extragalactic background light, early reionization, and x-rays~\cite{Arias:2012az}.  Constraints on the photon coupling from CAST and cooling of HB stars are weaker than the Xenon100 limits (see Ref.~\cite{Hochberg:2016ajh}), and are not shown here.

We see that semiconductor targets can probe sub-keV pseudoscalar DM, providing a strong alternative to model-dependent stellar constraints.

%%%%%%%%%%%%%%%%%%
\section{Conclusions}
In this note we proposed semiconductor targets, such as germanium and silicon, as detectors for light bosonic DM via an absorption process. We considered  electron excitation signals from absorption of halo DM with mass in the eV-keV range, as well as sub-eV DM emitted from the sun. Furthermore, DM in the few to 100~meV mass range can be absorbed and probed by these same targets via a two-phonon excitation process, if the sensitivity to phonon energy depositions is improved substantially.  We considered the reach in semiconductors for absorption of kinetically mixed hidden photons and pseudoscalars, demonstrating the strength of these targets to sub-keV DM. We also showed that current CDMSlite and DAMIC results already start to probe new parameter space, while future experiments such as SuperCDMS SNOLAB and DAMIC100 can cover a mass range that is currently wide open.

The two-phonon excitation studied here for DM absorption could also be utilized for probing DM in the keV to MeV mass range via scattering in semiconductors, in line with the two-excitation scattering in superfluid helium proposed in Ref.~\cite{Schutz:2016tid}.  We leave study of such scattering for future work.

%%%%%%%%%%%%%%%%%%
\vspace{0.2cm}
{\it Note added:} While completing this work, we became aware of Ref.~\cite{Bloch:2016sjj} which considers related topics.
\vspace{0.2cm}

%%%%%%%%%%%%%%%%%%
\vspace{0.2cm}
\mysections{Acknowledgments}
We thank Ritoban Basu Thakur, Alvaro Chavarria, Alan Robinson, and Matt Pyle for useful discussions. YH is supported by the U.S. National Science Foundation under Grant No. PHY-1002399.  KZ and TL are supported by the DoE under contract DE-AC02-05CH11231. TL is also supported by NSF grant PHY-1316783.

%%%%%%%%%%%%%%%%%%%%%%%%%%%%%%%%%%%%%%%%

%\bibliographystyle{utphys}
\bibliography{abssemi}

\end{document}